\documentclass[10pt,conference]{IEEEtran}
\usepackage{cite}
\usepackage{amsmath,amssymb,amsfonts}
\usepackage{graphicx}
\usepackage{textcomp}
\usepackage{xcolor}
\usepackage{booktabs}
\usepackage{xspace}
\usepackage{listings}
\usepackage{url}
\usepackage{tikz}
\usetikzlibrary{positioning,arrows.meta}
\usepackage[hidelinks]{hyperref}

\newcommand{\vs}{\textsc{VulSpot}\xspace}

\newcommand{\code}[1]{\texttt{\small #1}}

\lstdefinestyle{vs}{
  language=C, basicstyle=\ttfamily\scriptsize, columns=fullflexible, keepspaces=true,
  frame=tb, framerule=0.3pt, escapeinside={(*@}{@*)}, showstringspaces=false,
  keywordstyle=\bfseries, aboveskip=2pt, belowskip=2pt}

\begin{document}

\title{Beyond the Diff — \vs: A Vulnerability Dataset Where No Vulnerability Goes Unlocalized}

\author{
\IEEEauthorblockN{Dyna Soumhane Ouchebara}

  \IEEEauthorblockA{
    \textit{University of Mons}\\
    Mons, Belgium \\
    dynasoumhane.ouchebara@umons.ac.be}
\and
\IEEEauthorblockN{St\'ephane Dupont}

  \IEEEauthorblockA{
    \textit{University of Mons}\\
    Mons, Belgium \\
    stephane.dupont@umons.ac.be}
}

\maketitle

\begin{abstract}
  A vulnerability detector is only as trustworthy as the benchmark that scores it, and today's real-world datasets tend to flatter detectors. They are often built once and never refreshed, so they quickly miss new vulnerabilities, their labels are either noisy or heavily filtered, and their random splits let models train on near-copies of test code and on future vulnerabilities. Above all, when a fix only adds code, such as a missing NULL check, they do not point at where the vulnerability was, preventing detectors from learning to localize some of the most common C/C++ weaknesses. We present \vs, a living benchmark for detecting, classifying, and localizing vulnerabilities, with over 500k functions in 21 languages from 24k registered CVEs, including over 15k vulnerable C/C++ functions. By linking the code added by a patch to the existing statement it protects, \vs locates the vulnerability in 99.3\% of the functions that previous diff-based datasets used to leave without a location. Rather than discarding uncertain labels, it tells users how reliable each one is, and it comes with leak-free splits, complementary code context, and hierarchical CWE labels. An automated pipeline refreshes \vs every month, so that each release brings vulnerabilities no model has seen yet.
  \end{abstract}

\begin{IEEEkeywords}
vulnerability dataset, vulnerability detection, vulnerability localization, CWE classification
\end{IEEEkeywords}

\section{Introduction}
\label{sec:intro}
Learning-based vulnerability detectors are expected to say \emph{whether} a function is vulnerable, \emph{which} weakness (CWE, for Common Weakness Enumeration) it contains, and \emph{where} it is located, but how far we can trust their results depends on their training and evaluation data.

Most real-world datasets follow the same recipe: link CVEs (Common Vulnerabilities and Exposures) to their fixing commits, label each changed function as vulnerable before the fix and safe after it, and take the lines deleted or modified by the fix as vulnerable. BigVul~\cite{fan2020bigvul} and CVEfixes~\cite{bhandari2021cvefixes} applied this at scale, DiverseVul~\cite{chen2023diversevul} and MegaVul~\cite{ni2024megavul} increased size and code representations, PrimeVul~\cite{ding2025primevul} made labels stricter, and SecVulEval~\cite{ahmed2025secvuleval} added statement labels and LLM-generated context (Table~\ref{tab:related}). The recipe scales well, but it has issues that make reported results less reliable than they seem.

\begin{table}[t]
\centering
\caption{Comparison with real-world vulnerability datasets. Until = last CVE year; Loc.\ = statement- or line-level labels; Add.\ = additive fixes localized; Conf.\ = per-label confidence; Clone = clone-aware splits; Time = chronological split; Ctx.\ = context beyond the function; Hier.\ = hierarchical CWE labels; Upd.\ = automatic updates.}
\label{tab:related}
\footnotesize
\setlength{\tabcolsep}{2pt}
\resizebox{\columnwidth}{!}{%
\begin{tabular}{lrccccccccc}
\toprule
Dataset & Vuln.\ fn. & Until & Loc. & Add. & Conf. & Clone & Time & Ctx. & Hier. & Upd. \\
\midrule
BigVul~\cite{fan2020bigvul}          & 10,900$^\star$ & 2019 & \checkmark & -- & -- & -- & -- & -- & -- & -- \\
DiverseVul~\cite{chen2023diversevul} & 18,945$^\star$ & $\le$2023 & -- & -- & -- & -- & -- & -- & -- & -- \\
MegaVul~\cite{ni2024megavul}         & 20,408$^\dagger$ & 2023 & \checkmark & -- & -- & -- & -- & -- & -- & $\circ$ \\
PrimeVul~\cite{ding2025primevul}     & 6,968$^\star$  & $\le$2023 & -- & -- & -- & -- & \checkmark & -- & -- & -- \\
CleanVul~\cite{li2024cleanvul}       & 8,198$^\dagger$ & $\le$2025 & -- & -- & LLM & -- & -- & -- & -- & -- \\
SecVulEval~\cite{ahmed2025secvuleval} & 10,998$^\star$ & 2024 & \checkmark & -- & -- & -- & -- & LLM & -- & -- \\
TitanVul~\cite{li2026titanvul}       & 38,548$^\dagger$ & 2024 & -- & -- & -- & -- & \checkmark & -- & -- & -- \\
\vs \code{cpp}                       & 15,932$^\star$ & 2026 & \checkmark & \checkmark & Rule & \checkmark & \checkmark & AST & \checkmark & \checkmark \\
\vs \code{multi}                     & 28,334$^\dagger$ & 2026 & \checkmark & \checkmark & Rule & \checkmark & \checkmark & AST & \checkmark & \checkmark \\
\bottomrule
\end{tabular}}
\\[1pt]\scriptsize $^\dagger$Multilingual (CleanVul/MegaVul are C/C++ and Java). $^\star$C/C++ only. $\le$: cutoff not reported, bounded by the collection date. $\circ$: Crawler released for manual re-run.
\vspace{-3mm}
\end{table}

\textbf{Frozen datasets.} Datasets frozen at release soon miss recent vulnerabilities: the latest datasets stop in 2025 at the latest~\cite{li2024cleanvul}, and recent publications may inherit old cutoffs (TitanVul~\cite{li2026titanvul} aggregates seven datasets that stop in 2024). Such datasets also cannot evaluate LLMs on vulnerabilities disclosed after the models' training cutoff, which is the simplest way to verify if a model has already seen the test data (contamination)~\cite{sainz2023contamination}.

\textbf{Noisy labels caused by tangled commits.} A fixing commit often mixes the security fix with unrelated changes~\cite{herzig2013tangled,croft2023dataquality}, so the recipe labels non-vulnerable functions as vulnerable: manual audits found only 25\%, 52\%, and 60\% of correct vulnerable labels in BigVul, CVEfixes, and DiverseVul, respectively~\cite{chen2023diversevul,ding2025primevul}. Three remedies exist, each with its own flaw. \emph{Manual curation} is the most accurate (94\% for SVEN~\cite{ding2025primevul}), but small: SVEN \cite{he2023sven} (1,606 functions, 9 CWEs), BenchVul~\cite{li2026titanvul} (50 functions per Top-25 CWE, 82\% LLM-synthesized), and CASTLE~\cite{dubniczky2025castle} (250 hand-written C programs) can evaluate a model, but not train one. \emph{Strict filtering} (PrimeVul~\cite{ding2025primevul}) keeps only functions that are the single change of their commit or named in the CVE description, raising accuracy to 86--92\% but discarding all other true vulnerabilities. \emph{LLM validation} (CleanVul~\cite{li2024cleanvul}, 90.6\%; TitanVul~\cite{li2026titanvul}, 94\%) ties labels to costly, non-deterministic LLMs. All three either keep or drop labels, imposing a precision--coverage trade-off on users.

\textbf{Empty locations for additive fixes.} The statement-/line-level labels used by localizers, are usually defined as the lines deleted or modified by the fix~\cite{fan2020bigvul,ni2024megavul,ahmed2025secvuleval}. However, \emph{a fix that only adds code has no such line}, so the vulnerable function gets an empty location: it is often silently dropped from datasets or worse, counted as having no vulnerable line ~\cite{fu2022linevul}. Some localizers attempted to address this by labeling every line control- or data-dependent on the inserted code~\cite{li2021ivdetect,hin2022linevd} as vulnerable, but these were never recorded as such in datasets. These additive fixes, ex. missing checks, are the usual root cause of NULL-pointer dereferences, out-of-bounds accesses, and memory leaks: the most frequent C/C++ weaknesses, yet, to our knowledge, no dataset clearly localizes them.

\textbf{Leaky splits.} Under random splits, a model can train on: a function that is a near-duplicate of a test function (frequently its own fixed version) \cite{allamanis2019duplication}, and on vulnerabilities disclosed after those they are tested on~\cite{jimenez2019labelling}. We call this data leakage. PrimeVul~\cite{ding2025primevul} introduced chronological splits but near-duplicates still cross them, so the problem is but partly solved.

\textbf{The Linux kernel shift.} Since becoming a CVE Numbering Authority in February 2024, the Linux kernel has assigned thousands of CVEs per year~\cite{kernelcve}. Existing datasets predate or ignore this shift, but a dataset collected today easily becomes a Linux benchmark: until October 2026, 75.1\% of vulnerable C/C++ functions come from the kernel, mostly after 2024.

\textbf{Missing context and coarse CWE labels.} Models often fail because the needed information lives outside the function (a callee, a macro, a type definition), and LLM-generated contexts (SecVulEval~\cite{ahmed2025secvuleval}) are costly to regenerate and prone to model artifacts. Moreover, raw CWEs form a long tail of hundreds of classes, and benchmarks that shorten it impose one granularity, such as the Top-25 CWEs~\cite{cwetop25} of BenchVul~\cite{li2026titanvul} or the 8 OWASP categories~\cite{owasptop10} of SecLens~\cite{halder2026seclens}.

To address these issues, we present \vs, a dataset for function-level vulnerability detection, CWE classification, and statement-level localization, with 503,332 functions in 21 languages . Each contribution answers one of the issues above (data and code: \href{https://huggingface.co/datasets/ods01/VulSpot}{\textcolor{blue}{\underline{Hugging Face}}},
 \href{https://github.com/DynaSoumhaneOuchebara/VulSpot-Dataset}{\textcolor{blue}{\underline{GitHub}}}):
\begin{itemize}
    \setlength{\itemsep}{0pt}
    \item \textbf{A living dataset}: CVEs until October 2026 and a resumable pipeline that updates \vs every month each release being a new contamination-free test window.
    \item \textbf{Reliability-aware labels}: instead of a hard filter, a \emph{confidence tier} for every vulnerable label and a \emph{provenance type} for every negative, computed without LLMs.
    \item \textbf{Localization of additive fixes}: \emph{sink candidates} map each group of inserted statements to the statement of the vulnerable version it is tied to by data flow, locating 99.3\% of the vulnerabilities diff-based labels leave empty.
    \item \textbf{Leakage-aware evaluation and control of the Linux shift}: clone-aware chronological and project-holdout splits, high-confidence, paired and non-Linux test sets, and a balanced variant that caps post-2024 kernel CVEs.
    \item \textbf{Richer annotations}: deterministic AST-retrieved context and four hierarchical CWE views.
\end{itemize}

\section{Dataset Construction}
\label{sec:construction}

\begin{figure}[t]
  \centering
  \resizebox{\columnwidth}{!}{%
  \begin{tikzpicture}[
    box/.style={draw, rounded corners=1.5pt, fill=#1, align=center,
      font=\fontsize{6}{7}\selectfont, text width=1.8cm,
      minimum height=1.12cm, inner sep=1.5pt},
    arr/.style={-{Stealth[length=3pt]}, semithick},
    node distance=0.3cm]
  \node[box=blue!8] (n1) {\textbf{1. CVEs}\\NVD API 2.0, all non-rejected CVEs until the cutoff};
  \node[box=blue!8, right=of n1] (n2) {\textbf{2. Patch links}\\\texttt{Patch} references to GitHub and kernel.org};
  \node[box=blue!8, right=of n2] (n3) {\textbf{3. Commits}\\message, diff, files; no merges, backports, multi-commit CVEs};
  \node[box=orange!14, right=of n3] (n4) {\textbf{4. Functions}\\tree-sitter parsing and before/after pairing};
  \node[box=orange!14, below=0.42cm of n4] (n5) {\textbf{5. Labels}\\statement edit scripts, sink candidates, confidence tiers, negative types};
  \node[box=orange!14, left=of n5] (n6) {\textbf{6. Enrichment}\\deduplication, clone groups, AST context, CWE views};
  \node[box=green!12, left=of n6] (n7) {\textbf{7. Release}\\C/C++ and multi versions, full/balanced/paired variants, chrono.\ and project splits};
  \node[box=green!12, left=of n7] (n8) {\textbf{8. Upload}\\Parquet files on Hugging Face, versioned cutoff};
  \foreach \a/\b in {n1/n2,n2/n3,n3/n4,n4/n5,n5/n6,n6/n7,n7/n8} \draw[arr] (\a) -- (\b);
  \draw[arr, dashed, teal!70!black] (n8.west) -- ++(-0.3,0) |- (n1.west)
    node[pos=0.25, left, rotate=90, anchor=south, font=\fontsize{5.5}{6}\selectfont] {every month};
  \end{tikzpicture}}
  \vspace{-5mm}
  \caption{The \vs dataset construction pipeline.}
  \label{fig:pipeline}
  \vspace{-3mm}
  \end{figure}
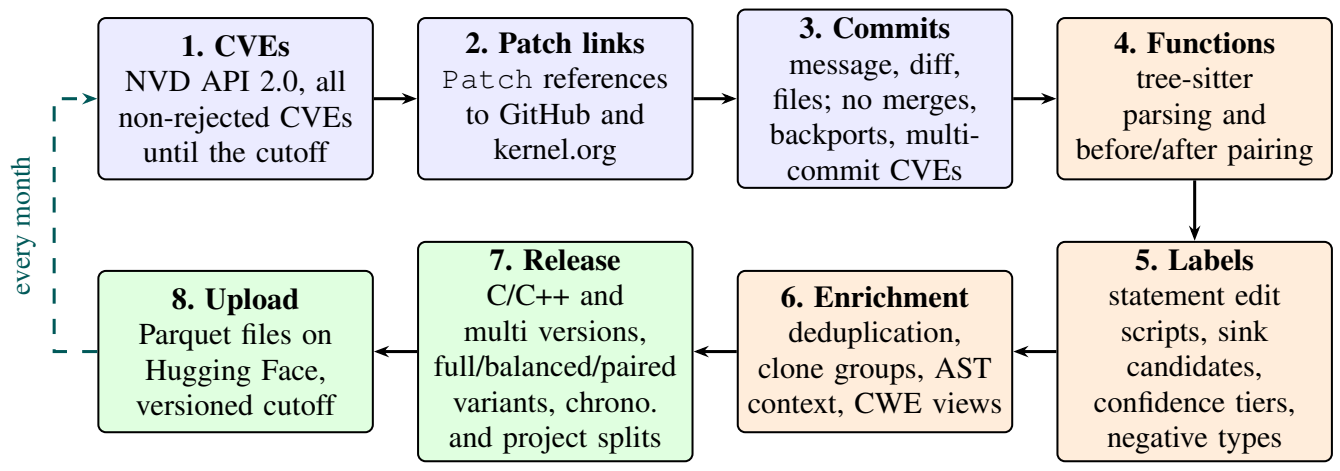

\subsection{Collection}
\textbf{CVEs and patches.} Following the pipeline of Fig.~\ref{fig:pipeline}, we download all 404,454 CVE records from the NVD API~2.0~\cite{nvdapi} and keep the non-rejected ones. Like the standard recipe, we only follow references tagged \code{Patch}: 30,773 CVEs point to a commit on GitHub or \code{git.kernel.org}; fork commits are attributed to the parent repository.

\textbf{Commits.} For each commit, we retrieve the message, the diff, and each modified file before and after the change, dropping merge commits, commits with over 300 files, and files over 2\,MB. NVD lists each kernel fix with its stable backports, so we detect them and drop those whose CVEs are covered by a mainline commit. Remaining multi-commit CVEs are dropped.

\textbf{Functions.} Files are parsed with tree-sitter~\cite{treesitter} and functions are paired across versions by name, signature, and order. A changed function yields a vulnerable (before) and a fixed (after) row; one removed by the fix is vulnerable, one added by it is negative. Of 1,251,421 extracted rows, 503,332 remain after filtering and exact deduplication.

\subsection{Statement-level labels}
\textbf{Statement edit scripts.} Each function is segmented into \emph{statement units}: simple statements, the headers of compound statements (\code{if (len > max)}), tails such as \code{while (c);}, the signature, and comments. Units are aligned across versions, guided by the file diff, and each receives an action: unchanged, updated, deleted, inserted, or moved. Whitespace-only edits are considered as "unchanged", and comment edits do not count as code edits. We also store deleted and updated statements with file-absolute line numbers for line-level labels.

\begin{figure}[t]
\begin{lstlisting}[style=vs]
data->strval = (char *) malloc(sizeof(char)*(size+1));
(*@\textcolor{teal}{+ if (!data->strval) \{ plist\_free\_data(data); ...; return NULL; \}}@*)
(*@\colorbox{orange!25}{\ttfamily\scriptsize memcpy(data->strval, *bnode, size);}@*) (*@\textcolor{orange!70!black}{$\leftarrow$ sink}@*)
data->strval[size] = '\0';
\end{lstlisting}
\vspace{-2mm}
\caption{Snippet of CVE-2017-6435. The fix only adds a NULL check (green); \vs anchors it on the \texttt{memcpy} that uses the unchecked pointer.}
\label{fig:example}
\vspace{-3mm}
\end{figure}

\textbf{Sink candidates.} Sink candidates are the statements a localizer should point at, starting with those deleted and updated by the fix. Inserted statements do not exist in the vulnerable version, so we group consecutive insertions and map each group to an \emph{anchor}: the statement of the vulnerable version tied to the inserted code by a shared variable (Fig.~\ref{fig:example}). \emph{Data-flow} rules follow the variables of the group to find the right statement to flag. They are grouped by how specific their pattern is: \emph{strong} (a guard and the use it protects, a release and its acquisition, a lock region, a status check and the call that set it, an initialization and its use, reorderings), \emph{medium} (relaxed guard--use, enclosing jump), and \emph{weak} (a shared identifier only). When no statement shares an identifier with the group, we fall back to its position: the next statement, or the previous one at the end of the function. Reordering fixes are anchored too: a statement moved later is its own anchor, as it ran too early (e.g., a release before the last use), and one moved earlier anchors on the first statement it jumped over that uses what it assigns (e.g., a copy that ran before its length was clamped). Each candidate stores its kind, rule, lines, byte range, and text, so users can choose diff-only or anchor-inclusive targets, or keep only the strongest rules.

\begin{table}[t]
\centering
\caption{Statistics of the six versions of \vs. Confidence tiers (high / medium/  low) are computed on vulnerable functions, negative types (untouched / fixed / added by the fix / patched-later-vulnerable) on safe functions.}
\label{tab:stats}
\footnotesize
\setlength{\tabcolsep}{2.5pt}
\resizebox{\columnwidth}{!}{
\begin{tabular}{lrrrrrr}
\toprule
& \multicolumn{3}{c}{\code{cpp}} & \multicolumn{3}{c}{\code{multi}} \\
\cmidrule(lr){2-4}\cmidrule(lr){5-7}
& full & balanced & paired & full & balanced & paired \\
\midrule
All functions             & 283,361 & 107,977 & 29,782 & 503,332 & 327,759 & 53,198 \\
Vulnerable functions            & 15,932  & 5,629   & 14,891 & 28,334  & 18,025  & 26,599 \\
\quad share (\%)      & 5.6     & 5.2     & 50.0   & 5.6     & 5.5     & 50.0 \\
CVEs                  & 16,176  & 5,332   & 12,079 & 24,257  & 13,407  & 18,270 \\
Repositories          & 753     & 753     & 688    & 3,634   & 3,634   & 3,112 \\
Distinct CWEs         & 231     & 217     & 228    & 429     & 420     & 426 \\
Linux share (vuln., \%) & 75.1  & 29.7    & 74.3   & 42.3    & 9.3     & 41.6 \\
Tiers h/m/l (\%)      & 81/17/3 & 67/27/7 & 80/17/3 & 61/36/3 & 45/50/5 & 61/36/3 \\
Neg.\ u/f/a/p (\%)    & 93/6/.4/.6 & 94/5/.4/.6 & 0/91/0/9 & 92/6/1.1/.5 & 92/6/1.5/.4 & 0/93/0/7 \\
Additive fixes (vuln., \%) & 32.6 & 32.4 & 33.2 & 25.4 & 21.2 & 26.0 \\
Empty diff labels (vuln., \%) & 27.3 & 27.8 & 27.8 & 23.5 & 21.5 & 24.1 \\
\bottomrule
\end{tabular}}
\vspace{-3mm}
\end{table}
\begin{table}[t]
\centering
\caption{Localization profile of frequent CWEs (\code{cpp/full}, grouped by CWE-1003): vulnerable functions fixed only by insertions (Add.) and with a data-flow anchor (DF anchors).}
\label{tab:cwe}
\small
\setlength{\tabcolsep}{3pt}
\begin{tabular}{llrrr}
\toprule
CWE & Weakness & $n$ & Add. & DF anchors \\
\midrule
476 & NULL dereference     & 2,467 & 39.6\% & 56.6\% \\
416 & Use after free       & 1,787 & 29.1\% & 52.2\% \\
125 & Out-of-bounds read   & 1,195 & 29.5\% & 48.9\% \\
787 & Out-of-bounds write  & 1,183 & 29.3\% & 50.0\% \\
401 & Memory leak          & 1,132 & 42.3\% & 65.8\% \\
362 & Race condition       & 589   & 26.5\% & 48.6\% \\
190 & Integer overflow     & 453   & 18.5\% & 36.2\% \\
\bottomrule
\end{tabular}
\vspace{-3mm}
\end{table}

\subsection{Label reliability}
\textbf{Confidence tiers.} Instead of discarding functions from tangled commits, each vulnerable function gets a confidence tier saying how sure we are that the CVE fixes it: \emph{high} if the commit changes a single function or it is named in the CVE description; \emph{medium} if it is the only changed function of its file in a multi-file commit; \emph{low} if its file has several changed functions and it is named in the commit message. Other changed functions of multi-function files, and test or fuzzing files are never labeled vulnerable.

\textbf{Provenance-aware negatives.} A \code{neg\_type} column marks negatives (safe functions) as \emph{fixed} versions, \emph{untouched} functions of the same files (up to 20 per file), functions \emph{added by the fix}, and \emph{patched-later-vulnerable} functions (i.e., fixed versions that a later CVE patches again). This helps debugging, as difficulty differs: a fixed function differs from its vulnerable version by a median of two statements, while an untouched one may be unrelated, thus easier to label. It also lets users restrict negatives to fixed versions, since untouched functions are only assumed safe and may hide unknown vulnerabilities.

A \code{clean} flag combines all quality signals: it is false for low-tier labels, label conflicts, patched-later-vulnerable negatives, and vulnerable functions whose "fix" only changes comments.

\subsection{Enrichment and release}
\textbf{Duplicates.} Exact duplicates (whitespace-insensitive MD5) are removed, keeping the earliest occurrence by CVE publication date. When the copies of a body carry different labels, the kept row is flagged with \code{label\_conflict} (4,682 rows in \code{multi}), typically because a fix was later reverted or an unfixed vulnerable function was copied elsewhere. As exact deduplication misses near-duplicates (renamed variable or a vendored copy), we link functions that share at least 80\% of their 5-token sequences (Jaccard similarity, approximated with MinHash-LSH~\cite{broder1997minhash}) and give each resulting group a \code{clone\_group\_id}: in \code{cpp}, 46,708 rows belong to 20,864 groups of two or more near-duplicates.

\textbf{Context.} From the AST and file, we extract each function's callees (with same-file definitions), types, macros, and globals.

\textbf{CWE views.} Instead of keeping only the Top-k CWEs, which drops much of the data, we offer CWE views, a rather unexplored granularity. By walking up the CWE hierarchy to the nearest member of each view, we map the original CWE of each row onto the right category in the following views: CWE-1000 (Research Concepts)~\cite{cwe}, CWE-1003~\cite{cwe1003} (Simplified Mapping of Vulnerabilities), CWE-699~\cite{cwe699} (Software Development) and CWE-1400~\cite{cwe1400} (Software Assurance Trends). These views give smaller, easier-to-learn class sets that cover most (if not all) of the data and allow per-category analysis.

\textbf{Variants, splits and evaluation subsets.} \vs exists in two configurations, \code{cpp} (C/C++ only) and \code{multi} (all 21 languages), each in three variants: \code{full} (all functions), \code{paired} (vulnerable functions and their fixed versions as safe functions), and \code{balanced}, which caps post-2024 kernel CVEs per calendar year at the kernel's average yearly CVE count before 2024. For each, we suggest two data splits. In the \textit{chronological} split, commits are sorted by the earliest publication date of their CVEs and assigned 80/10/10 to train, validation, and test (stratified by CWE), keeping all rows of a commit together. A row whose clone group already appears in an earlier split is excluded from the later one, so no test function is a near-duplicate of a training function and training never contains data newer than the test set. The \textit{project holdout} split keeps randomly chosen 10\% whole projects for testing. We also provide evaluation subsets: \code{test\_high} (high-confidence vulnerable functions), \code{test\_paired\_high} (these and their fixed versions, for paired evaluation), and \code{test\_no\_linux} (non-Linux functions).

\textbf{Automatic update.} Every month, a scheduled job collects CVEs published since the last cutoff, reruns all stages, uploads the new version, and schedules the next one.

\section{Dataset Statistics}
\label{sec:dataset}

\textbf{Overview.} \vs is distributed on Hugging Face as Parquet files, with code copied as-is under its original licenses. Table~\ref{tab:stats} summarizes its six variants. CVEs span 2009 to October 2026. Vulnerable functions are 5.6\% of \code{full}, a realistic imbalance, while \code{paired} contains 14,891 C/C++ vulnerable/fixed pairs. The Linux shift is visible: the kernel provides 75.1\% of vulnerable C/C++ functions in \code{full}, and 29.7\% in \code{balanced}. Positive labels are mostly high confidence (81\% in \code{cpp/full}; \code{multi} relies more on the medium tier, 36\%), and most negatives are untouched functions (93\%).

\textbf{Localization.} Fixes are small: they delete, update, or insert a median of two statements (7.8\% of a function), so knowing \emph{where} matters. Diff-based labels fail on additive fixes in two ways. In \code{cpp/full}, 5,194 vulnerable functions (32.6\%) are \emph{additive}: their fix inserts code but deletes or modifies no code statement. For 4,248 (27.3\%) of them, the label is \emph{empty}: the line diff marks no line of the vulnerable version. For the other 946 (5.9\% of vulnerable functions), the label is \emph{wrong}: the diff marks lines, but only because the inserted code re-indents them (e.g., an existing statement wrapped in a new \code{if} block), adds a brace to them, changes a comment, or because the diff aligns the insertion with the wrong copy of an identical line. Sink candidates avoid both failures, since they are computed on the statement edit script, which ignores whitespace and comments, rather than on the line diff: they localize 99.3\% of the functions with empty labels, and all 946 functions with noisy labels, whose candidates are data-flow anchors only (quality ones, not the less reliable positional fallbacks). Of all sink candidates, 70.9\% are deleted or updated statements (which diff-based datasets also provide) and 29.1\% are anchors (which only \vs provides). \emph{Strong} rules give 12.8\% of all candidates, \emph{medium} rules 4.9\%, \emph{weak} rules 7.2\%, and positional fallbacks 4.3\%. For functions with empty diff labels, the most reliable anchor is strong for 54.5\%, medium for 8.0\%, weak for 27.7\%, and positional for only 7.6\%. Table~\ref{tab:cwe} shows that additive fixes are most frequent for memory leaks (42.3\%), where the fix inserts a missing \code{free}, and NULL-pointer dereferences (39.6\%), where it inserts a check. Dropping them would thus bias localization benchmarks against some of the most common C/C++ weaknesses.

\textbf{CWE views.} 95\% of the 231 raw C/C++ CWEs map onto CWE-1000 and CWE-1400 views (these views covering all non-deprecated CWEs by design, the 5\% unmapped ones are only deprecated NVD categories), 84\% onto CWE-1003, and 71\% onto CWE-699 (these views covering only the most frequent software CWEs). These views shorten the long tail without dropping data: indeed, the 10 most frequent CWE-1400 categories cover 97\% of CWE-labeled vulnerable C/C++ functions, against 75\% for the 10 most frequent raw CWEs (90\% versus 53\% in \code{multi}, even more striking).

\textbf{Splits.} The chronological split of \code{cpp/full} has 12,487 / 1,482 / 1,548 vulnerable functions in train / validation / test, after the clone rule excluded 4,517 rows from later splits; the project holdout tests on 1,635 vulnerable functions from unseen projects. The test split yields 1,260 high-tier, 1,174 paired high-tier, and 337 non-Linux vulnerable functions. Excluding near-duplicates matters: under a random 80/10/10 row split, 57\% of vulnerable test functions in \code{cpp/full} have a near-duplicate in training (50\% their own fixed version), and under a random commit split, still 16\%, against 0\% in our splits, which justifies the value of our strategy.

\section{Research Opportunities}
\label{sec:opportunities}
Why yet another dataset? The issues mentioned in Section~\ref{sec:intro} systematically inflate reported results. \vs controls or exposes each, so its effect can be measured instead of silently inherited. This makes it useful in many use-cases:

\textbf{(1) Multiple tasks and granularities.} The same rows support four tasks that can be studied separately or learned jointly: \emph{function-level binary detection}, with the natural class imbalance or on the paired subset, where a model must tell a vulnerable function from a safe one or its own fix; \emph{CWE classification} at any granularity, from the 10 pillars of CWE-1000 to hundreds of raw CWEs, with nested views enabling hierarchical classifiers and milder penalties for errors close in the hierarchy; \emph{statement- and line-level localization}, now including additive fixes, which also enables checking whether detectors are right for the right sinks; and \emph{vulnerability repair}, on 14,891 C/C++ vulnerable/fixed pairs with edit scripts.

\textbf{(2) Context-aware detection.} The context columns allow controlled studies of how much and which kind of outside information (callees, macros, types, globals) a detector needs.

\textbf{(3) Learning with noisy labels.} Confidence tiers, negative types, and label conflict declarations make label noise measurable, enabling studies of its impact and of remedies (sample weighting, positive-unlabeled learning...).

\textbf{(4) Generalization and contamination-free evaluation.} The project-holdout and chronological splits measure generalization to unseen projects and future vulnerabilities, and each data release allows for testing LLMs past their training cutoff and shows if detectors degrade over time. Moreover,\code{ multi} enables transferability studies across 21 languages (e.g., from C/C++ to memory-safe languages and low-resource ones).

\section{Conclusion and Future Improvements}
\label{sec:conclusion}
We presented \vs, a living dataset of 503,332 functions that localizes additive-fix vulnerabilities, measures the reliability of its labels, and is split without leakage, to evaluate whether detectors can find, name, and locate unseen vulnerabilities. Despite its advantages, it still has some limitations: confidence tiers quantify and reduce, but do not remove, label noise (a single-function commit can be a partial fix, and untouched negatives may hide unknown vulnerabilities), and anchors follow data flow only through shared identifiers. Future work comprises manually auditing each tier and anchor rule, multi-commit fixes, as well as given and missing CWE labels (for which we are blindly trusting the NVD for now). We also plan to publish per-release baselines, to make \vs a continuously updated vulnerability detection leaderboard.

\section*{Acknowledgment}
This study was funded by CyberExcellence project by Digital Wallonia, and benefited from computational resources made available on Lucia, the Tier-1 supercomputer of the Walloon Region, infrastructure funded by the Walloon Region under the grant agreement n\textsuperscript{o}1910247.

\bibliographystyle{IEEEtran}
\bibliography{refs}

@inproceedings{fan2020bigvul,
  author    = {Jiahao Fan and Yi Li and Shaohua Wang and Tien N. Nguyen},
  title     = {A {C/C++} Code Vulnerability Dataset with Code Changes and {CVE} Summaries},
  booktitle = {Proc. Int. Conf. on Mining Software Repositories (MSR)},
  pages     = {508--512},
  year      = {2020}
}

@inproceedings{chen2023diversevul,
  author    = {Yizheng Chen and Zhoujie Ding and Lamya Alowain and Xinyun Chen and David Wagner},
  title     = {{DiverseVul}: A New Vulnerable Source Code Dataset for Deep Learning Based Vulnerability Detection},
  booktitle = {Proc. Int. Symp. on Research in Attacks, Intrusions and Defenses (RAID)},
  year      = {2023}
}

@inproceedings{ding2025primevul,
  author    = {Yangruibo Ding and Yanjun Fu and Omniyyah Ibrahim and Chawin Sitawarin and Xinyun Chen and Basel Alomair and David Wagner and Baishakhi Ray and Yizheng Chen},
  title     = {Vulnerability Detection with Code Language Models: How Far Are We?},
  booktitle = {Proc. IEEE/ACM Int. Conf. on Software Engineering (ICSE)},
  year      = {2025}
}

@inproceedings{bhandari2021cvefixes,
  author    = {Guru Bhandari and Amara Naseer and Leon Moonen},
  title     = {{CVEfixes}: Automated Collection of Vulnerabilities and Their Fixes from Open-Source Software},
  booktitle = {Proc. Int. Conf. on Predictive Models and Data Analytics in Software Engineering (PROMISE)},
  year      = {2021}
}

@inproceedings{ni2024megavul,
  author    = {Chao Ni and Liyu Shen and Xiaohu Yang and Yan Zhu and Shaohua Wang},
  title     = {{MegaVul}: A {C/C++} Vulnerability Dataset with Comprehensive Code Representations},
  booktitle = {Proc. Int. Conf. on Mining Software Repositories (MSR)},
  year      = {2024}
}

@inproceedings{he2023sven,
  author    = {Jingxuan He and Martin Vechev},
  title     = {Large Language Models for Code: Security Hardening and Adversarial Testing},
  booktitle = {Proc. ACM Conf. on Computer and Communications Security (CCS)},
  year      = {2023}
}

@misc{ahmed2025secvuleval,
  author       = {Md Basim Uddin Ahmed and Nima Shiri Harzevili and Jiho Shin and Hung Viet Pham and Song Wang},
  title        = {{SecVulEval}: Benchmarking {LLMs} for Real-World {C/C++} Vulnerability Detection},
  howpublished = {arXiv preprint arXiv:2505.19828},
  year         = {2025}
}

@inproceedings{li2026titanvul,
  author    = {Yikun Li and Ngoc Tan Bui and Ting Zhang and others},
  title     = {Out of Distribution, Out of Luck: How Well Can {LLMs} Trained on Vulnerability Datasets Detect Top 25 {CWE} Weaknesses?},
  booktitle = {Proc. IEEE/ACM Int. Conf. on Software Engineering (ICSE)},
  pages     = {2096--2108},
  year      = {2026},
  doi       = {10.1145/3744916.3787830}
}

@misc{li2024cleanvul,
  author       = {Yikun Li and Ting Zhang and Ratnadira Widyasari and others},
  title        = {{CleanVul}: Automatic Function-Level Vulnerability Detection in Code Commits Using {LLM} Heuristics},
  howpublished = {arXiv preprint arXiv:2411.17274},
  year         = {2024}
}

@misc{dubniczky2025castle,
  author       = {Richard A. Dubniczky and Krisztofer Zolt{\'a}n Horv{\'a}t and Tam{\'a}s Bisztray and Mohamed Amine Ferrag and Lucas C. Cordeiro and Norbert Tihanyi},
  title        = {{CASTLE}: Benchmarking Dataset for Static Code Analyzers and {LLMs} towards {CWE} Detection},
  howpublished = {arXiv preprint arXiv:2503.09433},
  year         = {2025}
}

@misc{halder2026seclens,
  author       = {Subho Halder and Siddharth Saxena and Kashinath Kadaba Shrish and Thiyagarajan M},
  title        = {{SecLens}: Role-Specific Evaluation of {LLMs} for Security Vulnerability Detection},
  howpublished = {arXiv preprint arXiv:2604.01637},
  year         = {2026}
}

@inproceedings{croft2023dataquality,
  author    = {Roland Croft and M. Ali Babar and M. Mehdi Kholoosi},
  title     = {Data Quality for Software Vulnerability Datasets},
  booktitle = {Proc. IEEE/ACM Int. Conf. on Software Engineering (ICSE)},
  year      = {2023}
}

@inproceedings{jimenez2019labelling,
  author    = {Matthieu Jimenez and Renaud Rwemalika and Mike Papadakis and Federica Sarro and Yves {Le Traon} and Mark Harman},
  title     = {The Importance of Accounting for Real-World Labelling When Predicting Software Vulnerabilities},
  booktitle = {Proc. ACM Joint European Software Engineering Conf. and Symp. on the Foundations of Software Engineering (ESEC/FSE)},
  year      = {2019}
}

@inproceedings{allamanis2019duplication,
  author    = {Miltiadis Allamanis},
  title     = {The Adverse Effects of Code Duplication in Machine Learning Models of Code},
  booktitle = {Proc. ACM SIGPLAN Int. Symp. on New Ideas, New Paradigms, and Reflections on Programming and Software (Onward!)},
  year      = {2019}
}

@inproceedings{herzig2013tangled,
  author    = {Kim Herzig and Andreas Zeller},
  title     = {The Impact of Tangled Code Changes},
  booktitle = {Proc. Working Conf. on Mining Software Repositories (MSR)},
  year      = {2013}
}

@inproceedings{fu2022linevul,
  author    = {Michael Fu and Chakkrit Tantithamthavorn},
  title     = {{LineVul}: A Transformer-Based Line-Level Vulnerability Prediction},
  booktitle = {Proc. Int. Conf. on Mining Software Repositories (MSR)},
  year      = {2022}
}

@inproceedings{hin2022linevd,
  author    = {David Hin and Andrey Kan and Huaming Chen and M. Ali Babar},
  title     = {{LineVD}: Statement-Level Vulnerability Detection Using Graph Neural Networks},
  booktitle = {Proc. Int. Conf. on Mining Software Repositories (MSR)},
  year      = {2022}
}

@inproceedings{li2021ivdetect,
  author    = {Yi Li and Shaohua Wang and Tien N. Nguyen},
  title     = {Vulnerability Detection with Fine-Grained Interpretations},
  booktitle = {Proc. ACM Joint European Software Engineering Conf. and Symp. on the Foundations of Software Engineering (ESEC/FSE)},
  year      = {2021}
}

@inproceedings{broder1997minhash,
  author    = {Andrei Z. Broder},
  title     = {On the Resemblance and Containment of Documents},
  booktitle = {Proc. Compression and Complexity of Sequences},
  year      = {1997}
}

@inproceedings{sainz2023contamination,
  author    = {Oscar Sainz and Jon Ander Campos and Iker Garc{\'\i}a-Ferrero and Julen Etxaniz and Oier Lopez de Lacalle and Eneko Agirre},
  title     = {{NLP} Evaluation in Trouble: On the Need to Measure {LLM} Data Contamination for Each Benchmark},
  booktitle = {Findings of the Association for Computational Linguistics: EMNLP},
  year      = {2023}
}

@misc{nvdapi,
  author       = {{National Institute of Standards and Technology}},
  title        = {{NVD} Vulnerability {API} 2.0},
  howpublished = {\url{https://nvd.nist.gov/developers/vulnerabilities}},
  note         = {Accessed 2026}
}

@misc{cwe1003,
  author       = {{MITRE}},
  title        = {{CWE} View 1003: Weaknesses for Simplified Mapping of Published Vulnerabilities},
  howpublished = {\url{https://cwe.mitre.org/data/definitions/1003.html}},
  note         = {Accessed 2026}
}

@misc{kernelcve,
  author       = {{The Linux Kernel Documentation}},
  title        = {{CVEs}},
  howpublished = {\url{https://docs.kernel.org/process/cve.html}},
  note         = {Accessed 2026}
}

@misc{treesitter,
  author       = {Max Brunsfeld and others},
  title        = {Tree-sitter: An Incremental Parsing System for Programming Tools},
  howpublished = {\url{https://tree-sitter.github.io/}},
  note         = {Accessed 2026}
}

@misc{cwe699,
  author       = {{MITRE}},
  title        = {{CWE} View 699: Software Development},
  howpublished = {\url{https://cwe.mitre.org/data/definitions/699.html}},
  note         = {Accessed 2026}
}

@misc{cwe1400,
  author       = {{MITRE}},
  title        = {{CWE} View 1400: Comprehensive Categorization for Software Assurance Trends},
  howpublished = {\url{https://cwe.mitre.org/data/definitions/1400.html}},
  note         = {Accessed 2026}
}

@misc{cwe,
  author       = {{MITRE}},
  title        = {Common Weakness Enumeration ({CWE})},
  howpublished = {\url{https://cwe.mitre.org/}},
  note         = {Accessed 2026}
}

@misc{cwetop25,
  author       = {{MITRE}},
  title        = {{CWE} Top 25 Most Dangerous Software Weaknesses},
  howpublished = {\url{https://cwe.mitre.org/top25/}},
  note         = {Accessed 2026}
}

@misc{owasptop10,
  author       = {{OWASP Foundation}},
  title        = {{OWASP} Top Ten},
  howpublished = {\url{https://owasp.org/www-project-top-ten/}},
  note         = {Accessed 2026}
}

\end{document}